\documentclass[
 reprint,
 superscriptaddress,
 amsmath,amssymb,
 aps,
]{revtex4-2}

\usepackage{graphicx}
\usepackage{dcolumn}
\usepackage{bm}
\usepackage[colorlinks=true, citecolor=Blue, linkcolor=Navy, urlcolor=RoyalBlue]{hyperref}

\usepackage{multirow} 

\newcommand{\smallMatrix}[1]{\begin{pmatrix}#1\end{pmatrix}}
\newcolumntype{P}[1]{>{\centering\arraybackslash}p{#1}}
\newcolumntype{M}[1]{>{\centering\arraybackslash}m{#1}}

\usepackage{diagbox}
\usepackage[svgnames,table]{xcolor}

\usepackage{caption}
\usepackage{subcaption}
\usepackage{ragged2e}
\usepackage{mathrsfs}
\usepackage{braket}

\DeclareCaptionJustification{justified}{\justifying}

\renewcommand{\arraystretch}{1.6} 
\begin{document}


\title{Quantum noise reduction schemes for KAGRA post-O5 upgrade}

\author{Yuhang Zhao}
\email{yuhang@hnas.ac.cn}
\affiliation{Institute for Gravitational Wave Astronomy, Henan Academy of Sciences, Zhengzhou, 450046, China}

\author{Marc Eisenmann}
\address{National Astronomical Observatory of Japan, 2-21-1 Osawa, Mitaka, Tokyo, 181-8588, Japan}

\author{Michael Page}
\address{National Astronomical Observatory of Japan, 2-21-1 Osawa, Mitaka, Tokyo, 181-8588, Japan}

\author{Zong-Hong Zhu}
\email{zhuzh@bnu.edu.cn}
\affiliation{School of Physics and Astronomy, Beijing Normal University, Beijing 100875, China}

\affiliation{Institute for Frontiers in Astronomy and Astrophysics, Beijing Normal University, Beijing 102206, China}

\begin{abstract}
Quantum noise, arising from the quantization of electromagnetic field, has been a limiting noise source for current gravitational wave detectors. Squeezed vacuum modifies quantum fluctuations and has been routinely employed. To reduce quantum noise, the current solution is to combine squeezed vacuum with a hundred-meter-long detuned overcoupled optical cavity (filter cavity) to achieve frequency-dependent squeezing (FDS). The sensitivity to gravitational wave signals can be decomposed into a noise budget. Depending on the detector configuration, the contribution from noise sources other than quantum noise can be significant. In particular, suspension noise from multi‑stage pendulums is a key factor in quantum‑noise reduction design. In the context of KAGRA post-O5, we have compared quantum noise reduction schemes, including single-mode squeezing techniques -- namely, frequency-independent squeezing (FIS), FDS with a filter cavity (FC), FDS with an amplitude filter cavity (AFC), and FDS with a frequency-dependent beam splitter (FDBS) -- as well as a two-mode squeezing technique (EPR scheme). The FC scheme was found to outperform the AFC and FDBS schemes at all frequencies. It was found that FIS scheme gives the largest Binary Neutron Star (BNS) range when low frequency noise is dominated by classical noise, while the FC scheme gives the largest BNS range when low-frequency noise becomes dominated by quantum noise. Optimized filter cavity parameters could substantially improve the BNS range. This would allow at least 23\% increase in the detection rate for an 85\,m filter cavity, compared with using FIS scheme. Once a filter cavity is constructed with optimized parameters, refining its detuning can fully compensate for the variations in arm power (from half to full design value) and for different intracavity loss conditions. The EPR scheme performs best for the detection of heavy binary systems.
\end{abstract}

\maketitle


\section{Introduction\label{sec-intro}}

Since the first direct detection\cite{abbott2016observation}, gravitational-wave astronomy has evolved remarkably, realised through the LIGO\cite{ligo2015advanced}-Virgo\cite{acernese2015advanced}-KAGRA\cite{aso2013interferometer} (LVK) Collaboration, which has now completed four global observing runs and publicly catalogued over two hundred events. Results of LVK observations have been used to elucidate strong-field gravity\cite{abbott2021tests}, black-hole demographics\cite{ray2023nonparametric}, neutron-star equation of state\cite{abbott2018gw170817}, nuclesosynthesis, and cosmology\cite{ligo2017gravitational} -- with the milestones such as the confirmation of Hawking's area theorem\cite{isi2021testing}, strong evidence for compact objects within the purported lower mass gap\cite{abac2024observation}, hierarchical growth of black holes\cite{abac2025gw241011}, direct evidence for intermediate-mass black holes\cite{abbott2020gw190521}\cite{abac2025gw231123} standing out as paradigm-shifting cases. Nevertheless, the current detection horizon and event rate remain fundamentally constrained by quantum noise.

Quantum noise is one of the most fundamental limitations for the sensitivity of interferometric gravitational wave detectors\cite{caves1981quantum}. It arises from the quantum uncertainty in the amplitude and phase of the vacuum fluctuations coupled to the electromagnetic field used to probe the strain of passing gravitational waves. This noise consists of high-frequency shot noise and low-frequency radiation pressure noise, both of which originate from the vacuum fluctuations entering the output port of the detector. Owing to the optical response of the detector, radiation pressure noise dominates at low frequencies while shot noise dominates at high frequencies. The output field is in a coherent vacuum state, and the product of the uncertainties in these two quadratures obeys the Heisenberg uncertainty principle. By employing nonlinear optics to generate squeezed vacuum\cite{gao2024generation}, correlations are created between sidebands symmetric about the carrier frequency; homodyne detection then reveals that noise is reduced in one quadrature at the expense of the other.

To suppress quantum noise across the entire detection bandwidth, one must inject frequency-dependent squeezed vacuum whose squeezing angle rotates from $0^\circ$ (phase squeezing) at high frequencies to $90^\circ$ (amplitude squeezing) at low frequencies. The seminal work by Kimble et al. demonstrated this using frequency-independent squeezing in conjunction with two detuned, over-coupled filter cavities \cite{kimble2001conversion}. These cavities store the squeezed vacuum within their bandwidth and introduce a frequency-dependent differential phase delay due to detuning, effectively rotating the squeezing angle by $90^\circ$ at low frequencies by altering the phase relation between the correlated sidebands.

In practice, gravitational wave detectors employ a dual-recycling configuration, which yields a double-pole optical response. In principle, this requires two filter cavities to achieve the required squeezing rotation. However, when the interferometer is operated in the resonant sideband extraction (RSE) configuration, a single filter cavity suffices, as first shown by Khalili et al. \cite{khalili2007quantum}; consequently, a two-mirror linear cavity was proposed and implemented for LIGO \cite{evans2013realistic}. 

Besides the standard filter cavity scheme, several alternate implementations exist. The so-called amplitude filter cavity concept utilizes a detuned cavity as a high-pass filter for a phase-squeezed state \cite{corbitt2004optical}. At high frequencies, the squeezed quadrature is reflected toward the detector, while at low frequencies, the anti-squeezed quadrature is transmitted through the cavity and replaced by uncorrelated coherent vacuum entering from the reverse direction. Although this approach does not achieve full broadband quantum noise reduction, it can mitigate harmful anti-squeezing with much less stringent loss requirements than a standard frequency-dependent squeezing (FDS) filter cavity. A logical extension of this concept employs double squeezed inputs—one beam to reduce high-frequency noise and the other to reduce low-frequency noise—such that the filter cavity effectively acts as a frequency-dependent beam splitter \cite{khalili2009increasing}.

All the aforementioned schemes rely on single-mode squeezing, which correlates sidebands symmetric about the carrier. Recently, a two-mode (EPR) squeezing scheme was proposed to achieve frequency-dependent noise reduction without external filter cavities \cite{ma2017proposal}. In this configuration, two beams are generated at different frequencies: the signal beam is resonant for the interferometer which acquires the gravitational wave signal, while the idler beam is detuned so that it sees the interferometer as a filter cavity, thereby acquiring the frequency dependence. This is equivalent to squeezing joint quadrature measurements—for instance, the difference and sum of the amplitude and phase quadratures of the two beams—which establishes correlations akin to EPR entanglement. The squeezed joint quadrature is thus rotated, enabling broadband noise reduction in a manner similar to external filter cavities. While this EPR approach avoids the need for additional vacuum envelopes, it is more susceptible to optical loss and significantly complicates the required signal processing for the joint quadrature readout. The EPR techniques are also important for quantum information \cite{liu2025continuous}.

Finally, for a general detuned signal-recycling configuration (as proposed for the Einstein Telescope low frequency interferometer\cite{punturo2010einstein}), the double-pole phase response necessitates more complex rotation methods. These include: (i) two sequential filter cavities, (ii) a coupled filter cavity \cite{jones2020implications} \cite{ding2025performance}, (iii) one filter cavity combined with EPR squeezing \cite{peng2024approaches}, or (iv) a quantum teleportation scheme that extends EPR squeezing using a two-mode squeezed state and a one-mode squeezed state \cite{nishino2024frequency}.

At this moment, only the Kimble scheme of frequency dependent squeezing has been tested at the desired frequency region\cite{mcculler2020frequency}\cite{zhao2020frequency} and implemented\cite{ganapathy2023broadband}\cite{acernese2023frequency} for gravitational wave detectors. The use of frequency dependent squeezing in LIGO's O4 run increases detection rate by about 65\%, a substantial improvement over standard frequency-independent methods\cite{acernese2019increasing}\cite{tse2019quantum}.

The next-generation gravitational wave detectors such as Einstein Telescope\cite{punturo2010einstein} and Cosmic Explorer\cite{reitze2019cosmic} will be more limited by quantum noise through technique improvements, making quantum noise reduction even more crucial.

For KAGRA post O5 plan\cite{akutsu2025decadal}, the low frequency sensitivity is quite limited by suspension thermal noise and mirror thermal noise. Thus achieving an excellent reduction of quantum noise at low frequency may not be significant. The performance of frequency dependent squeezing depends heavily on the losses and the length of the filter cavity\cite{capocasa2018measurement}.Usually shorter filter cavity has worse performance. Recently, the dephasing effect and locking precision is found to be the main issue when the length of filter cavity becomes shorter\cite{mcculler2021ligo}. Since the FDBS scheme\cite{khalili2009increasing} and AFC scheme\cite{corbitt2004optical} feature to have the optical cavity operated on resonance, it may be advantageous when low frequency behavior is less significant. Previous work from Khalili\cite{khalili2010optimal} made an exhaustive comparison among different schemes, but the bandwidth of interferometer has been optimized in his work. In this work, we keep the bandwidth of interferometer invariant. EPR scheme features a reduction of quantum noise at all frequencies, so it may behave better when we compare it with a frequency dependent squeezing realized by a short filter cavity. We will also consider FIS scheme since it features to have the lowest propagation optical losses due to less Faraday isolator used. These five schemes are relatively more mature and considered in this work, compared with other more advanced schemes\cite{page2021gravitational}. It is worth mentioning that we focus on the HF configuration of KAGRA post-O5 upgrade. It features a high reflectivity of the signal extraction mirror, introducing a gain of sensitivity with a dip around 2 and 3\,kHz. The high-frequency\,(HF) configuration highlights a unique scientific opportunity for KAGRA in the BNS post-merger regime.

The structure of this paper is organized as follows: section\,\ref{sec:fdbs} presents a general field coupling model for an optical cavity with dissipation, and its geometry description; section\,\ref{sec:loss_depha} presents an empirical and theoretical losses behavior for optical cavities with different length; section\,\ref{sec:quantum_correla} presented particular case of quantum noise for KAGRA and also suspension and mirror thermal noise; section\,\ref{sec:qn_kagra} summarizes the formalisms for quantum noise reduction schemes; section\,\ref{sec:conclusion} shows the comparison of different quantum noise reduction schemes for KAGRA post-O5 upgrade in terms of astrophysical implications.

\section{Optical cavities for squeezed vacuum systems}
\label{sec:fdbs}

Optical cavities serve as essential building blocks in various quantum noise reduction schemes. In this section, we review the interaction of an optical field --- which, in the context of quantum noise reduction, will be a squeezed vacuum --- with such a cavity. In particular, we describe how optical losses couple into the system and present an updated empirical assessment of the loss performance of optical cavities, normalized by their geometric parameter.

\subsection{Field interaction with optical cavities}

\begin{figure}[htbp]
    \centering
    \includegraphics[width=0.99\linewidth]{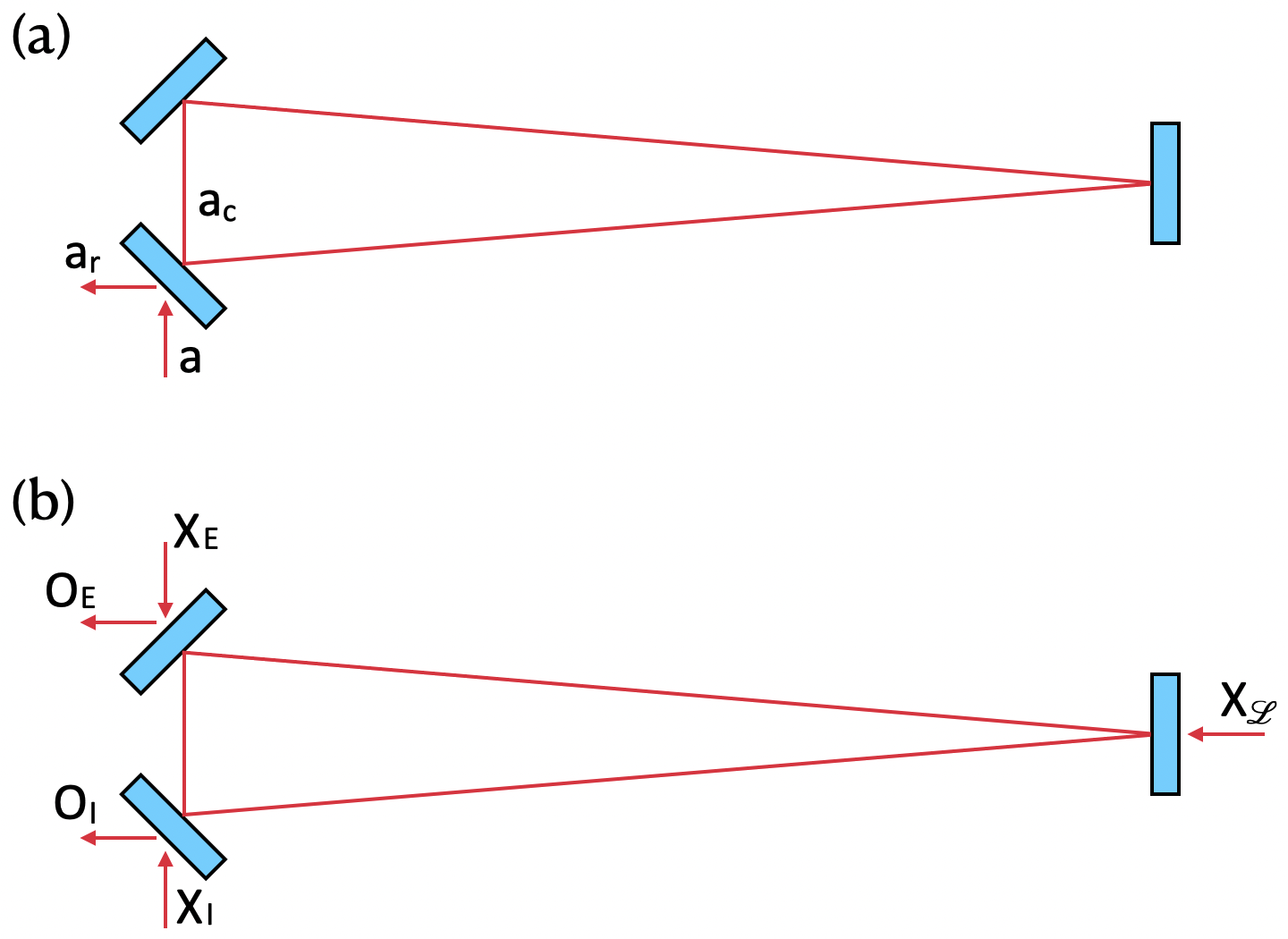}
    \caption{Schematic of an optical cavity. (a) Simplified configuration with a single input field $a$.  The field circulating inside the cavity is $a_c$, and $a_r$ denotes the field reflected from the input mirror. (b) Generalized configuration with three input fields -- $X_I$, $X_E$, and $X_\mathscr{L}$ -- injected through the input mirror, end mirror, and loss mirror, respectively. The reflectivity and transmissivity differ for each input-output port combination. Since the output from the loss port (right mirror) is inaccessible, only the output fields $O_I$ and $O_E$ leaving the input and end mirrors, respectively, are considered. }
    \label{fig：lossy_cavity}
\end{figure}

We can model the coupling of the optical field to the cavity using the scheme shown in Fig.\,\ref{fig：lossy_cavity} (a).The cavity reaches a steady state when the interference between the circulating field $a_c$ (after one round trip) and the input beam $a$ (injected through the input mirror) reproduces $a_c$ itself. This process can be formulated as in Ref. \cite{bond2016interferometer}
\begin{equation}
    a_c = i t_1 a + r_1 r_2 \sqrt{1-\mathscr{L}}a_c e^{-ikL},
\end{equation}

where $r_1$ and $t_1$ are the amplitude reflectivity and transmissivity of the input mirror, respectively, and $r_2$ is the amplitude reflectivity of the end mirror. Here, $\mathscr{L}$ denotes the round trip optical losses, $k$ is the wavenumber, and $L$ is the round trip length. Thus we obtain 
\begin{equation}
    a_c = a \frac{i t_1}{1-r_1 r_2 \sqrt{1-\mathscr{L}} e^{i kL}}
\end{equation}



If the cavity finesse exceeds 10\cite{slagmolen2000phase}, we can apply the approximation $r_1 = \sqrt{1-T_1} \approx 1-\frac{T_1}{2}$, where $T_1$ is the power transmissivity of the input mirror. Similarly, $r_2 = \sqrt{1-T_2} \approx 1-\frac{T_2}{2}$ and $\sqrt{1-\mathscr{L}} \approx 1-\frac{\mathscr{L}}{2}$. The phase accumulation per round trip is $e^{ikL} = e^{i\omega L/c} = e^{i(\omega+\Omega)\tau} \simeq e^{i\Omega\tau} \simeq 1 + i\Omega\tau$, where the first approximation assumes the carrier is on resonance ($e^{i\omega\tau}\simeq 1$) and the second holds for $\Omega\tau \ll 1$ (i.e., the sideband frequency is small compared with the cavity free spectral range). The round-trip time is $\tau = L/c$. With these approximations, the cavity reflection coefficient can be written as
\begin{equation}
\label{eqn:decay_rate_ref}
    a_r/a = \frac{\gamma_2+\gamma_{\mathscr{L}}-\gamma_1-i\Omega}{\gamma-i\Omega},
\end{equation}
where $\gamma_1 = {T_1}/{2\tau}$ is field decay rate due to the input mirror (equivalently, the half width half maximum in rad/s), and $\gamma_2$ and $\gamma_{\mathscr{L}}$ are the corresponding decay rates associated with the end mirror and intracavity losses, respectively. The total decay rate is $\gamma = \gamma_1 + \gamma_2 + \gamma_{\mathscr{L}}$.

To treat either the coupling of optical losses to squeezed states or the combination of squeezed states in an optical cavity, one must adopt the general three-input-field configuration depicted in Fig.\,\ref{fig：lossy_cavity}(b). The corresponding input-output relation is given by
\begin{equation}
\label{eqn:fc_full}
    \smallMatrix{O_I \\ O_E} = \smallMatrix{R_{II} & T_{EI} & T_{\mathscr{L}I} \\ T_{IE} & R_{EE} & T_{\mathscr{L}E}}\smallMatrix{X_I \\ X_E \\ X_{\mathscr{L}}},
\end{equation}

where $X_I$, $X_E$, and $X_{\mathscr{L}}$ are the fields injected through the input, end, and loss mirrors, respectively, and $O_I$ and $O_E$ are the fields exiting from the input and end mirrors, respectively. The $2\times 3$ matrix consists of the cavity reflection and transmission coefficients, with the subscripts indicating the input-output port pairs (e.g., $T_{EI}$ denotes transmission from the input-mirror input port to the end-mirror output port). In general, when these coefficients are neither 0 nor 1, each output field is a linear mixture of all three input fields. Since the coefficients are frequency-dependent, the cavity effectively behaves as a frequency-dependent beam splitter.

\subsection{Normalization of cavity geometry}

The geometry of a two-mirror linear cavity is determined by three parameters: the cavity length $\mathrm{L}$, and the radii of curvature $\mathrm{R_1}$ and $\mathrm{R_2}$ for input and end mirrors, respectively. The Rayleigh range is given by 
\begin{equation}
\label{eqn:rayleigh}
    z_0 = \sqrt{\frac{L(R_1-L)(R_2-L)(R_1+R_2-d)}{(R_1+R_2-2L)^2}}, 
\end{equation}
and the waist size is 
\begin{equation}
    \omega_0 = \sqrt{\frac{\lambda z_0}{\pi}}.
\end{equation}
For a symmetric confocal cavity, the two mirrors have equal radii of curvature, $R_1 = R_2 =L$, so their focal lengths are both $L/2$. Substituting $R=L+\epsilon$ into Eq.\,\ref{eqn:rayleigh} and taking the limit $\epsilon \rightarrow 0$, $z_0 = L/2$ can be obtained. Thus confocal cavity length satisfies
\begin{equation}
    L_{conf} = \frac{\pi \omega_0^2}{\lambda}.
\end{equation}
Since $L_{conf}$ is proportional to the waist size $\omega_0$, and the beam size on the mirrors scales with $\omega_0$, different cavities can be normalized to an effective confocal length based on the average beam size at the mirrors. This normalization has been adopted in previous works \cite{evans2013realistic}\cite{capocasa2018measurement}, as scattering losses are sensitive to the beam size -- the larger the beam size, the more significant the influence of the mirror surface profile on the loss.

\subsection{Empirical loss performance of optical cavities}
\label{sec:loss_depha}

Optical losses impose a fundamental constraint on the overall bandwidth of an optical cavity. As defined in Eq.\,\ref{eqn:decay_rate_ref}, losses-contributed bandwidth takes the explicit form

\begin{equation}
    \gamma_{loss} = \frac{\mathscr{L}}{2\tau} =\frac{c}{2}\frac{ \mathscr{L}}{L}.
\end{equation}
The loss-contributed bandwidth is governed by the loss per unit length. In practice, losses for quantum filter cavities are quoted per unit length\cite{evans2013realistic}. Fitting the aforementioned loss data against the equivalent confocal cavity length provides an empirical scaling law for the cavity performance
\begin{equation}
    \mathscr{L} = 10\times L_{conf}^{0.3}.
\end{equation}

\begin{figure}[htbp]
    \centering
    \includegraphics[width=1\linewidth]{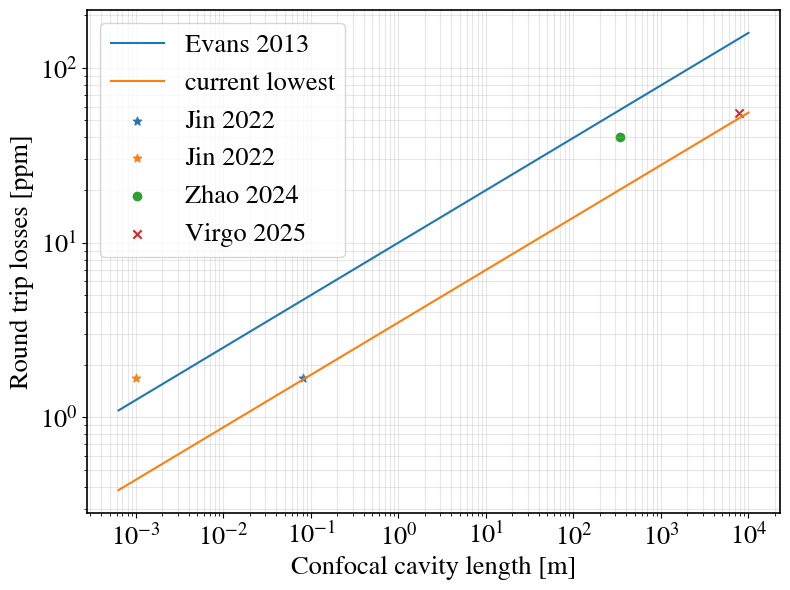}
    \caption{Optical losses as a fucntion of confocal cavity length. The blue curve is extrapolated from Evans et al\cite{evans2013realistic}. There are some measured optical losses from recent works such as micro-fabricated mirrors by Jin et al\cite{jin2022micro}, Virgo arm cavity mirrors\cite{virgo2025optical}, Virgo filter cavity mirrors by Zhao et al\cite{zhao2024optical}. Considering two points with lowest optical losses, a new line is plotted as "current lowest".}
    \label{fig:losses}
\end{figure}

Fig.\,\ref{fig:losses} plots the above relation as the blue curve, with the cavity geometry normalized by the confocal length . Selected recent measurements\cite{jin2022micro}\cite{virgo2025optical}\cite{zhao2024optical} are also shown for comparison. The figure also includes an empirical lower envelope representing the best reported loss performance, which is given by
\begin{equation}
    \mathscr{L}_{low} = 3.5\times L_{conf}^{0.3}.
\end{equation}

\begin{figure}[htbp]
    \centering
    \includegraphics[width=1\linewidth]{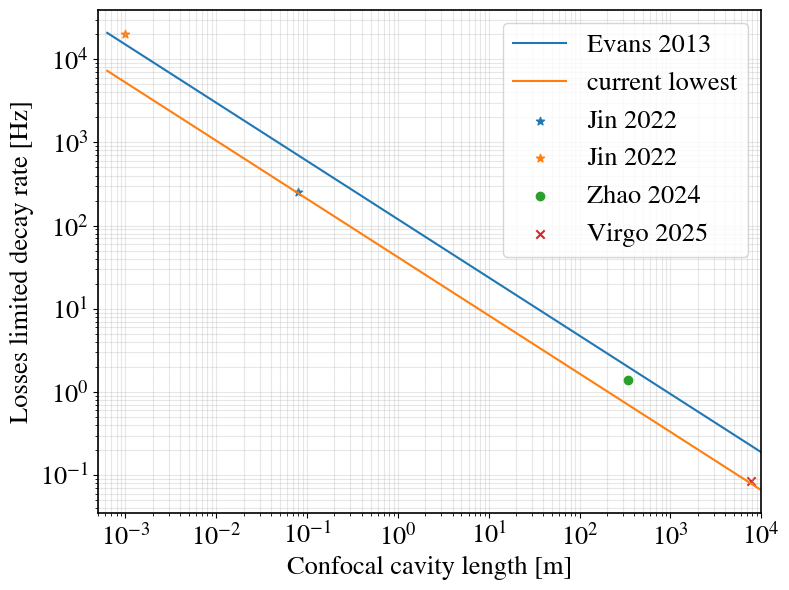}
    \caption{Contribution of optical losses to the total bandwidth as a fucntion of confocal cavity length.}
    \label{fig:losspermeter}
\end{figure}

The contribution of optical losses to the total bandwidth is illustrated in Fig.\,\ref{fig:losspermeter}. The figure shows that the impact of losses on the bandwidth becomes less significant as the cavity length increases. For the current-generation filter cavity of a few hundred meters, the corresponding loss-limited FWHM is on the order of a few hertz---roughly an order of magnitude smaller than the filter cavity bandwidth required by current interferometers \cite{ganapathy2023broadband}. For ETLF, where the filter cavity bandwidth is only a few hertz \cite{ding2025performance}, a cavity length of order kilometers is preferable to ensure that the loss-limited bandwidth remains a factor of 10 below the total filter cavity bandwidth.

The benefit of a long filter cavity is evident from Fig.\,\ref{fig:losspermeter}. A long cavity ensures that losses are negligible, thereby preserving the squeezed vacuum. Moreover, if state-of-the-art loss levels can be maintained, a significant reduction in the required cavity length would be achievable.

The dominant loss mechanisms are scattered light \cite{isogai2013loss} and higher-order modes excited by surface figure errors \cite{capocasa2016estimation}. Furthermore, mirror surface defects have been shown to introduce a spatial dependence to the optical losses \cite{zhao2024optical,kozlowski2025design}.

\section{Quantum noise of gravitational wave detectors}
\label{sec:quantum_correla}

\begin{figure}[htbp]
    \centering
    \includegraphics[width=1\linewidth]{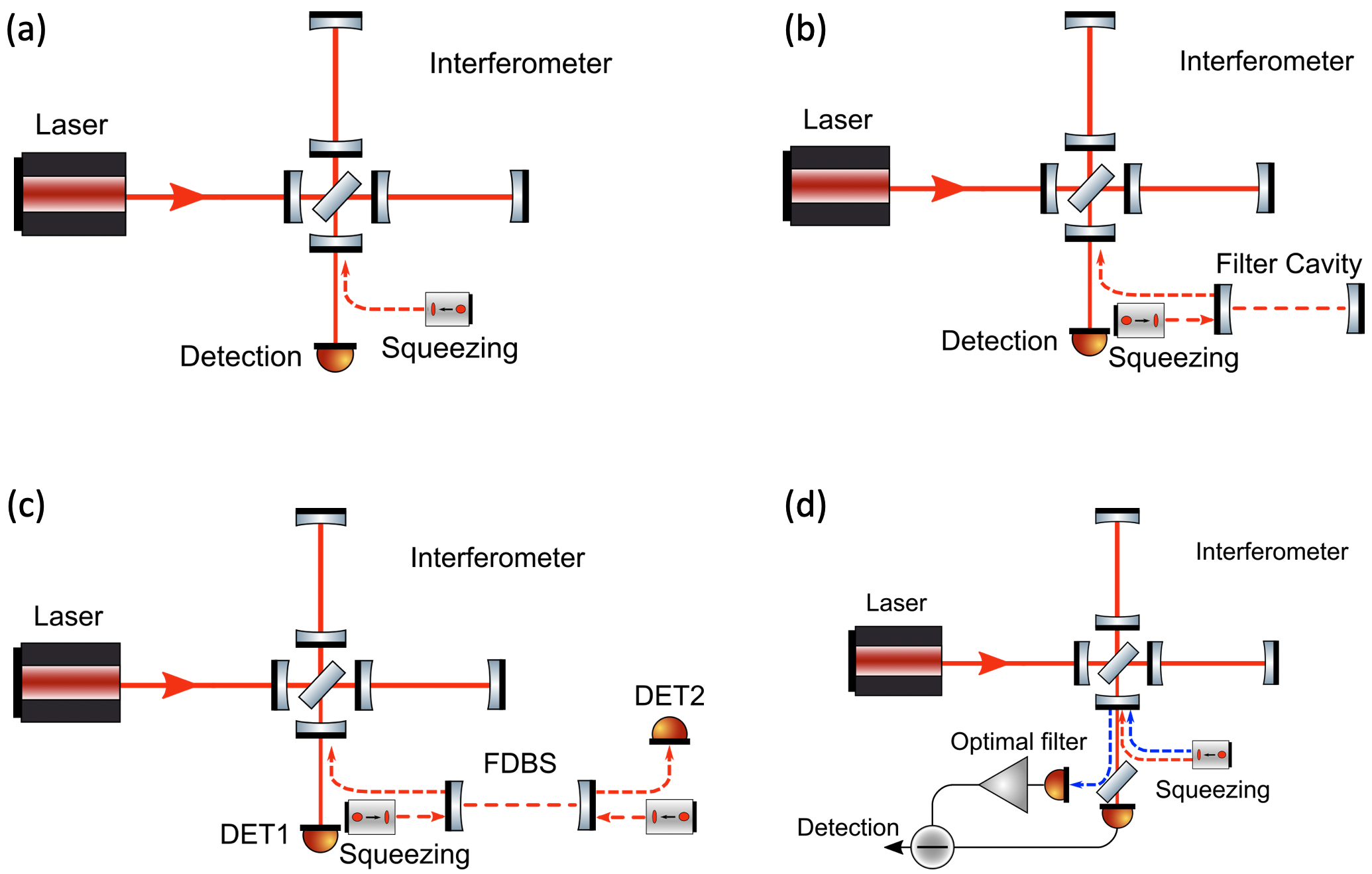}
    \caption{Schematic of quantum noise reduction schemes. (a) Frequency-independent squeezing (FIS); (b) filter cavity (FC) or amplitude filter cavity (AFC), depending on the mirrors transmissivity and cavity detuning; (c) frequency-dependent beam splitter (FDBS) with dual squeezed inputs; (d) Einstein-Podolsky-Rosen (EPR) squeezing scheme.}
    \label{fig:qn_schemes}
\end{figure}

Quantum noise in gravitational-wave detectors arises from vacuum fluctuations entering through various optical ports and ultimately reaching the readout port. The dominant contribution comes from the output port, which is why all quantum noise reduction schemes focus on modifying the vacuum field at this location, as illustrated in Fig.\,\ref{fig:qn_schemes}. The corresponding single-sided power spectral density of quantum noise, $S_h^{QN}$, is given by
\begin{equation}
\label{eqn:qn}
    S_h^{QN} = S_h^{in}+S_h^{arm}+S_h^{SR}+S_h^{vac},
\end{equation}
where $S_h^{in}$ denotes the spectral density of the injected vacuum entering the readout port, $S_h^{arm}$ accounts for losses in the interferometer arms, $S_h^{SR}$ accounts for losses in the signal-recycling cavity, and $S_h^{vac}$ accounts for losses associated with the detection process. A passing gravitational wave distorts spacetime and phase-modulates the sidebands symmetrically about the carrier frequency. In the two-photon formalism, each term in Eq.\,\ref{eqn:qn} represents the ratio of the readout quadrature magnitude arising from quantum fluctuations to that induced by the gravitational-wave signal. A detailed quantum noise budget for squeezed-vacuum injection can be found in Ref.~\cite{ding2025performance}.

The quantum noise reduction techniques investigated in this work all modify the injected-vacuum term of quantum noise, i.e., $S_h^{in}$ in Eq.\,\ref{eqn:qn} (see also Fig.\,\ref{fig:qn_schemes}). This term can be written explicitly as
\begin{equation}
\label{eqn:vac_in}
    S_h^{in} = \frac{u^T(T^{in}S^{in}T^{in\dagger})u}{|u^T{v^{h}}|^2},
\end{equation}
where $S^{in}$ denotes the spectral density of the injected vacuum, $T^{in}$ is the overall transfer function describing the interaction between the injected vacuum and the interferometer, $u$ represents the homodyne detection vector, and $v^{h}$ is the readout-port signal arising from the phase-modulated sidebands induced by gravitational waves in the arms, after their propagation through the interferometer.

For frequency-independent squeezing, the injected-vacuum spectral density can be written as
\begin{eqnarray}
\label{eqn:sqz}
    S^{in} &&= S(\sigma, \phi) \\
    &&=\smallMatrix{\cos\phi & -\sin\phi \\\sin\phi & \cos\phi}\smallMatrix{e^{\sigma} & 0 \\ 0 & e^{-\sigma}}\smallMatrix{\cos\phi & \sin\phi \\ -\sin\phi & \cos\phi} \nonumber
\end{eqnarray}

For the FC scheme, the injected-vacuum term is modified by the filter cavity transfer matrix $G$ as
\begin{equation}
\label{eqn:transfer_fc}
    S^{in} = GS(\sigma, \phi)G^\dagger,
\end{equation}
where G is the transfer matrix of filter cavity. However, to account for optical losses or the AFC/FDBS schemes, Eq.\,\ref{eqn:fc_full} must be considered and the $S^{in}$ term will become a sum of contribution from three different ports. The transfer matrix is first computed in the sideband formalism and then converted to the two-photon formalism. The sideband reflection coefficient of the filter cavity is given in Eq.\,\ref{eqn:fc_full}. Converting this expression to the two-photon formalism using the conversion matrix $A_2$ defined in Ref.~\cite{kwee2014decoherence},
\begin{equation}
    A_2 = \frac{1}{\sqrt{2}}\smallMatrix{1 & 1 \\ -i & +i},
\end{equation}
yields
\begin{equation}
    G = A_2\smallMatrix{r(\Omega) & 0 \\ 0 & r^*(-\Omega)}A_2^{-1}.
\end{equation}

The key difference between the FC and AFC schemes lies in the detuning and coupling condition of the filter cavity. For the FC scheme, the cavity is over-coupled, with minimal end-mirror transmission, and is operated with a detuning. For the AFC scheme, the cavity is critically coupled and operated at zero detuning. Although the same formal expression as in Eq.\,\ref{eqn:fc_full} applies to both configurations, the parameter regimes are entirely different.

For the FDBS scheme, the standard configuration of Eq.~\ref{eqn:fc_full}---where coherent vacuum only couples through the input mirror---is modified by injecting an additional squeezed vacuum through the end mirror. This second squeezed field is described by Eq.~\ref{eqn:sqz} but with a different squeezing angle $\phi$. The mixing of the two squeezed inputs $X_I$ and $X_E$ produces outputs $O_I$ and $O_E$ that exhibit quantum correlations. These correlations are preserved as one of the fields propagates through the interferometer [Fig.~\ref{fig:qn_schemes}(c)], thereby enabling a corresponding reduction of quantum noise. The outputs of DET1 (interferometer) and DET2 (filter cavity) are then combined with an optimal filter $K(\Omega)$ to yield the final output signal $o(\Omega)$ as follows:

\begin{equation}
    o(\Omega) = DET1(\Omega)-K(\Omega)DET2(\Omega).
\end{equation}
The power spectral density of the output signal is given by
\begin{eqnarray}
    S_{o}(\Omega) =&& S_{DET1}(\Omega)-2\Re[K(\Omega)S_{DET1,DET2}(\Omega)] \nonumber \\
    &&+|K(\Omega)|^2S_{DET2}(\Omega)
\end{eqnarray}

The optimal filter that minimizes the quantum noise is
\begin{equation}
    K(\Omega) = \frac{S_{\text{DET1,DET2}}(\Omega)}{S_{\text{DET1}}(\Omega)},
\end{equation}
which yields the minimum output noise spectrum
\begin{equation}
    S_o(\Omega) = S_h (\Omega) - \frac{S_{DET1,DET2}^2}{S_{DET1}}
\end{equation}
Note that the DET1 signal is the same as $S_h (\Omega)$, which has been used for the above equation.

The EPR scheme employs two-mode squeezing centered at two distinct frequencies: one resonant with the interferometer (the signal field) and the other detuned (the idler field). The signal field, resonant with the interferometer, obeys the same form as Eq.~\ref{eqn:vac_in}. The idler field, being detuned, sees the interferometer as a coupled filter cavity, whose transfer function takes the form of Eq.~\ref{eqn:transfer_fc}; it is then detected directly by a homodyne detector. The bandwidth of such a coupled cavity is given by
\begin{equation}
\label{eqn:epr_gamma}
    \gamma_{cc} = \frac{c\tau\tau^*}{4L},
\end{equation}
where $\tau$ is the amplitude transmissivity of the compound signal extraction cavity and can be written as
\begin{equation}
    \tau = \frac{i\sqrt{T_{SEM}T_{arm}}e^{i\phi_{SEC}}}{1-\sqrt{R_{SEM}R_{arm}}e^{2i\phi_{SEC}}}.
\end{equation}
The coupled cavity bandwidth will be determined by the interferometer standard quantum limit frequency. To achieve such bandwidth, the signal extraction cavity phase can be derived from Eq.\,\ref{eqn:epr_gamma} as
\begin{equation}
    \phi_{SEC} = \frac{1}{2}\arccos[\frac{1+R_{arm}R_{SEM}-\frac{\gamma_{arm}T_{SEM}}{\gamma_{cc}}}{2\sqrt{R_{arm}R_{SEM}}}]+n\pi
\end{equation}
The arm length and signal extraction cavity length are adjusted to be multiple times of the wavelength of carrier. Thus the values of these two length are obtained as shown in Tab.\,\ref{tab:FCpara}. The way of calculating the conditional squeezing is the same as the FDBS scheme.

\begin{table*}[h]
\caption{\label{tab:FCpara}Summary of the relevant interferometer parameters.}
\renewcommand{\arraystretch}{1.6}
\setlength{\tabcolsep}{6pt}
\begin{tabular}{lcr}
    \toprule
Parameter & Symbol & Value \\
\hline
Carrier wavelength & $\lambda$ & 1064 nm\\
Arm power & $\mathrm{P_{arm}}$ & 1.3 MW\\
Arm input mirror transmissivity & $\mathrm{T_{arm}}$ & 0.4 \%\\
Arm mirror mass & m & 40 kg \\
Signal extraction mirror transmissivity & $\mathrm{T_{SEM}}$ & 0.5 \% \\
Arm cavity length & $\mathrm{L_{arm}}$ & 2999.9999008560 m \\
Signal extraction cavity length & $\mathrm{L_{SEC}}$ & 66.6039283280 m \\
Idler shift frequency & $\Delta$ & 76.54739993 MHz \\
\hline
\multicolumn{3}{c}{\textsc{Optical losses budget}} \\
Injection optical losses & $l_{inj}$ & 4\% \\
Readout optical losses & $l_{ro}$ & 3\% \\
Signal extraction cavity losses & $l_{SEC}$ & 500\,ppm \\
\hline
\multicolumn{3}{c}{\textsc{Optimal filter cavity parameter}} \\
Input mirror transmissivity (L=40\,m, l=30\,ppm) & $T_{in}$ & 31\,ppm \\
Input mirror transmissivity (L=40\,m, l=10\,ppm) & $T_{in}$ & 26\,ppm \\
Input mirror transmissivity (L=85\,m, l=38\,ppm) & $T_{in}$ & 58\,ppm \\
Input mirror transmissivity (L=85\,m, l=13\,ppm) & $T_{in}$ & 63\,ppm \\
\hline
\end{tabular}
\end{table*}

\section{Quantum noise for KAGRA post O5}
\label{sec:qn_kagra}

The high-frequency (HF) configuration of KAGRA post-O5 will feature 1.3\,MW of laser power in each arm, higher quality factors for the sapphire fibers in the final suspension stage, larger test masses, and higher reflectivity of the signal-extraction mirror \cite{akutsu2025decadal}. For the quantum noise of the interferometer summarized in Table~\ref{tab:FCpara}, the frequency-dependent squeezing realized by a filter cavity requires a bandwidth (half-width at half-maximum, HWHM) and a detuning given by the following equations \cite{kwee2014decoherence}:
\begin{eqnarray}
    \gamma_{fc} &=& \sqrt{\frac{2}{(2-\epsilon)\sqrt{1-\epsilon}}}\frac{\Omega_{SQL}}{\sqrt{2}}, \\
    \delta_{fc} &=& \sqrt{1-\epsilon}\gamma_{fc},
    \label{eqn:fcpara}
\end{eqnarray}
where standard quantum limit frequency is
\begin{equation}
    \Omega_{SQL} = \frac{8}{c}\sqrt{\frac{P_{arm}\omega_0}{mT_{arm}}},
\end{equation}
and the factor associated to cavity losses is
\begin{equation}
    \epsilon = \frac{4}{2+\sqrt{2+2\sqrt{1+(\frac{2\Omega_{SQL}}{f_{FSR}l^2})^4}}}.
\end{equation}

For the parameters of the KAGRA post-O5 upgrade, the standard quantum limit frequency is 12.75\,Hz. However, as noted in Ref.~\cite{whittle2020optimal}, the above equation no longer yields the optimal filter-cavity parameters when losses contribute significantly to the cavity decay rate. In this work, we use particle swarm optimization to optimize three parameters: the squeezing level, the input-mirror transmissivity, and the detuning. The squeezing level considered in the optimization ranges from 10\,dB to 15\,dB, and 15\,dB of squeezing is consistently found to be optimal. This is mainly attributed to the low phase-noise level assumed in our simulation. To remain conservative, we therefore limit the injected squeezing to 15\,dB. The detuning, in contrast, is relatively straightforward to set to the desired value. The optimal input-mirror transmissivity depends on the cavity length and the round-trip losses, as summarized in Table~\ref{tab:FCpara}.

\begin{figure}[htbp]
    \centering
    \includegraphics[width=1\linewidth]{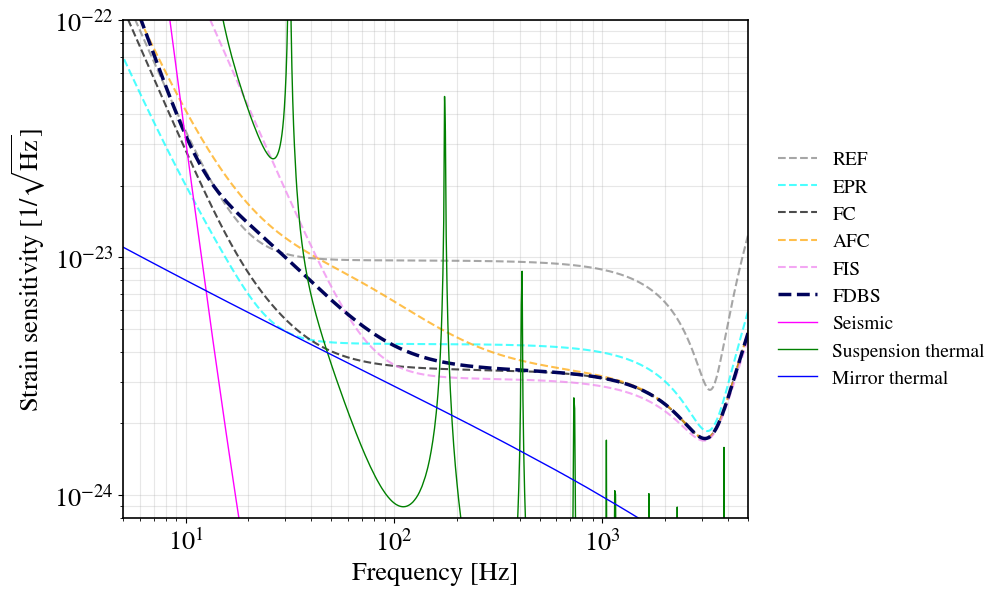}
    \caption{Quantum noise from different quantum noise reduction schemes and major classical noise curves. Here high quality suspension has been assumed to be used.}
    \label{fig:compare}
\end{figure}

The KAGRA interferometer, located at the Kamioka mine, provides approximately 85\,m of space for constructing a filter cavity. We assume 13\,ppm round-trip losses, which is a current-best state-of-art value. With these parameters, Fig.~\ref{fig:compare} compares the quantum noise reduction schemes. The EPR scheme offers the best low-frequency sensitivity, owing to the more effective squeezing-angle rotation achieved by using the interferometer itself as a filter cavity. The FIS scheme provides the best high-frequency sensitivity, though the difference between the FIS and FC schemes is small, as the main limitation arises from losses in the signal-extraction cavity. The FC scheme is optimal at intermediate frequencies. Neither the AFC nor the FDBS scheme outperforms the FC scheme, despite the absence of the dephasing effect associated with the latter.

The optimization of filter cavity parameters were done to optimize only BNS range. It should be noted that the optimal filter cavity parameters will be different if we chose to optimize for heavier astrophysical sources.

\section{Astrophysical implication \label{sec:conclusion}}

\begin{figure}[htbp]
    \centering
    \includegraphics[width=1\linewidth]{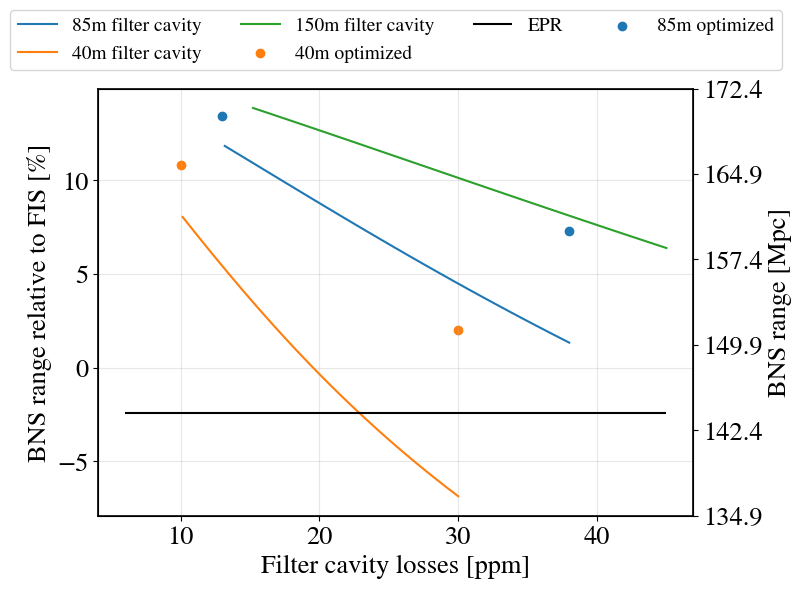}
    \caption{Detection range for KAGRA post-O5 after considering both quantum and classical noise (high quality factor suspension). The 85\,m space is current most allowed space in the KAGRA tunnel. All filter cavity losses value used here is between the best achievable value to the empirical line, as shown between the blue and orange line in Fig.\,\ref{fig:losses}. The optimization is done for the cases of 40\,m and 85\,m filter cavities for the lowest and highest filter cavity losses values.}
    \label{fig:range_compare}
\end{figure}

In this work, we have compared several quantum noise reduction schemes. Given the constraint on filter cavity length, we find trade-offs among the different schemes. To determine which scheme is most suitable for implementation in the KAGRA post-O5 upgrade, we evaluate the astrophysical reach, e.g., BNS range. Figure~\ref{fig:range_compare} compares the BNS range for the FIS, EPR, and FC schemes as a function of cavity length and optical losses. The 85\,m filter cavity outperforms the EPR scheme by 7\% to 14\%. This corresponds to a 23\% to 48\% increase in the detection rate.

\begin{figure}[htbp]
    \centering
    \includegraphics[width=1\linewidth]{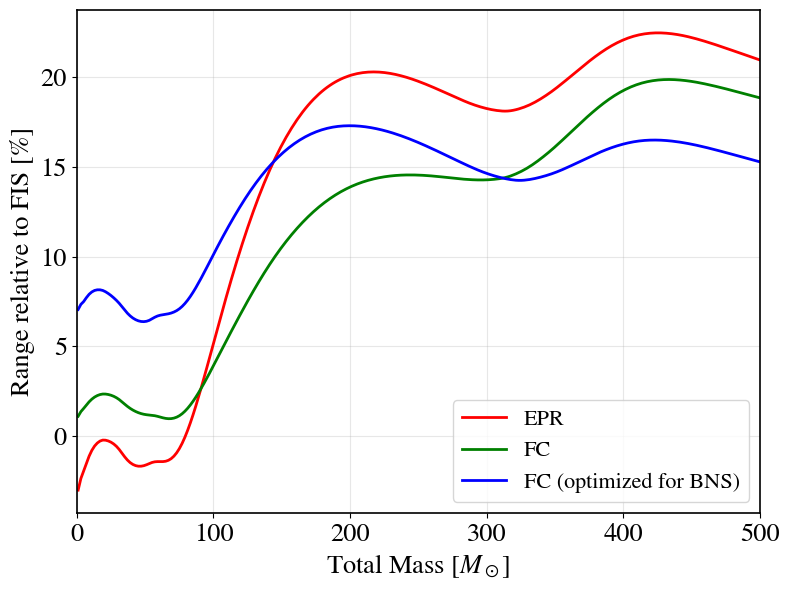}
    \caption{Detection range as a function of binary system total mass for KAGRA post-O5 by comparing different quantum noise reduction schemes. A 85\,m long filter cavity is assumed with 38\,ppm optical losses to be conservative. High quality suspension is assumed.}
    \label{fig:range_mass_compare}
\end{figure}

For an 85\,m filter cavity with 38\,ppm losses, Fig.~\ref{fig:range_mass_compare} compares the different quantum noise reduction schemes. The FC scheme performs best for lower-mass binary systems, whereas the EPR scheme is optimal for heavier systems. The figure also shows that while optimizing the filter cavity parameters increases the BNS range, a compromise is made for heavier binary systems.

\section{Conclusions}

\begin{figure}[htbp]
    \centering
    \includegraphics[width=1\linewidth]{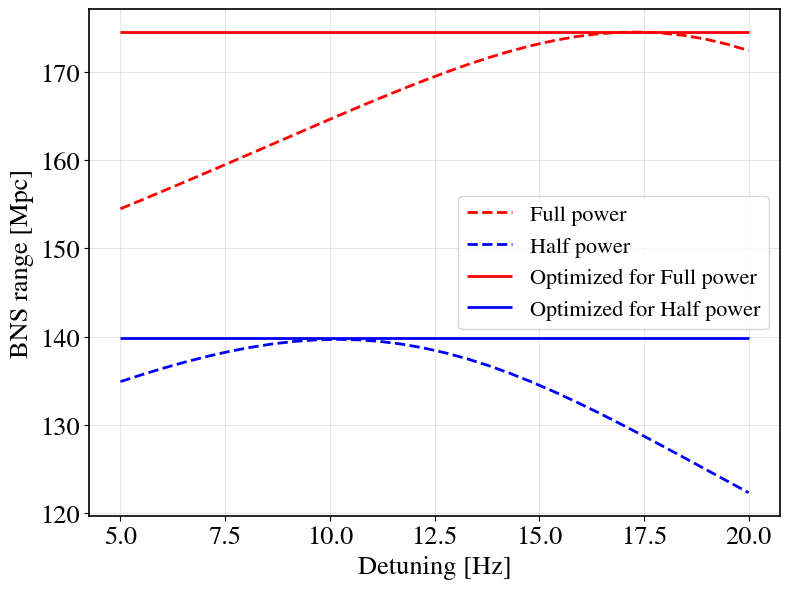}
    \caption{KAGRA post-O5 range when the detuning of a 85\,m filter cavity is varied. The optimized lines for full or half power are plotted using optimal input mirror transmissivity and cavity detuning for 13\,ppm intra-cavity losses. The dashed lines are BNS range with detuning varied. By varying detuning, the BNS range can be optimized to the best BNS range without changing input mirror transmissivity.}
    \label{fig:tune}
\end{figure}

Squeezing has become an indispensable technique for laser-interferometer gravitational-wave detectors. KAGRA's post-O5 upgrade stands to benefit from the implementation of such techniques. For an underground interferometer, the available space imposes practical constraints on the cavity length. For KAGRA, we have considered several quantum noise reduction techniques. The amplitude filter cavity (AFC) and frequency-dependent beam splitter (FDBS) schemes are not competitive. Frequency-independent squeezing is advantageous when the low-frequency band is dominated by classical noise. The conventional FC scheme, under the cavity-length constraint, offers the best BNS range but does not perform as well as the EPR scheme for heavier binary systems. It is also found that a numerical optimization of the input-mirror transmissivity is required to ensure optimal filter-cavity performance. As shown in Fig.~\ref{fig:range_compare}, the optimized points significantly outperform the results obtained from the analytical formula. For such loss-dominated cavities, variations in arm power or cavity losses would necessitate different input-mirror transmissivities. However, as shown in Fig.~\ref{fig:tune}, tuning the filter-cavity detuning can fully compensate for such variations.

\section{Appendix A: Another scenario of classical noise}

When suspension has low quality factor, the low frequency improvement from squeezing reduction would be negligible. Here we quantify such difference among different schemes.

\begin{figure}[htbp]
    \centering
    \includegraphics[width=1\linewidth]{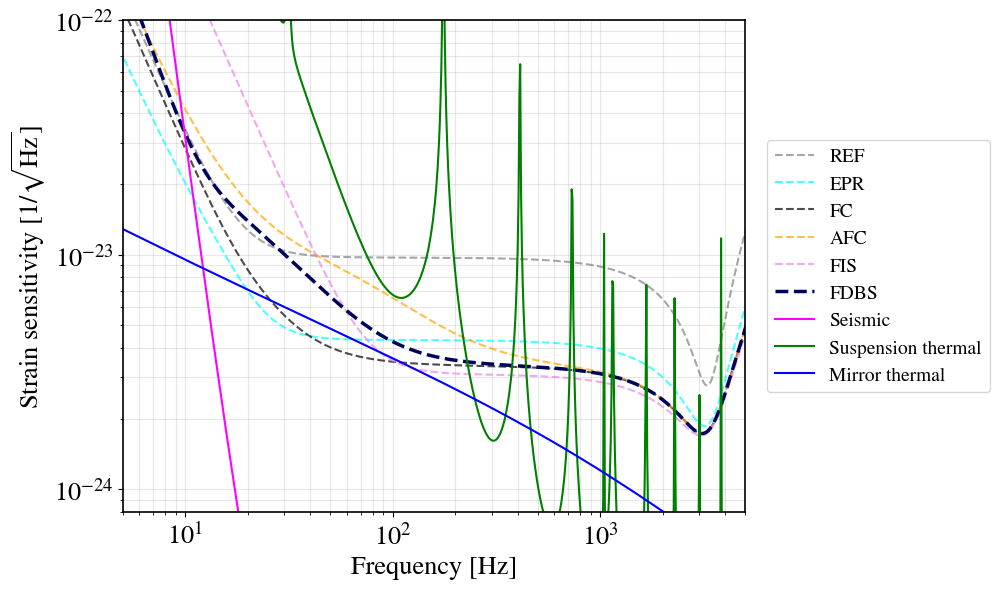}
    \caption{Quantum noise from different quantum noise reduction schemes and major classical noise curves. Here low quality suspension has been assumed to be used.}
    \label{fig:compare_LQS}
\end{figure}

\begin{figure}[htbp]
    \centering
    \includegraphics[width=1\linewidth]{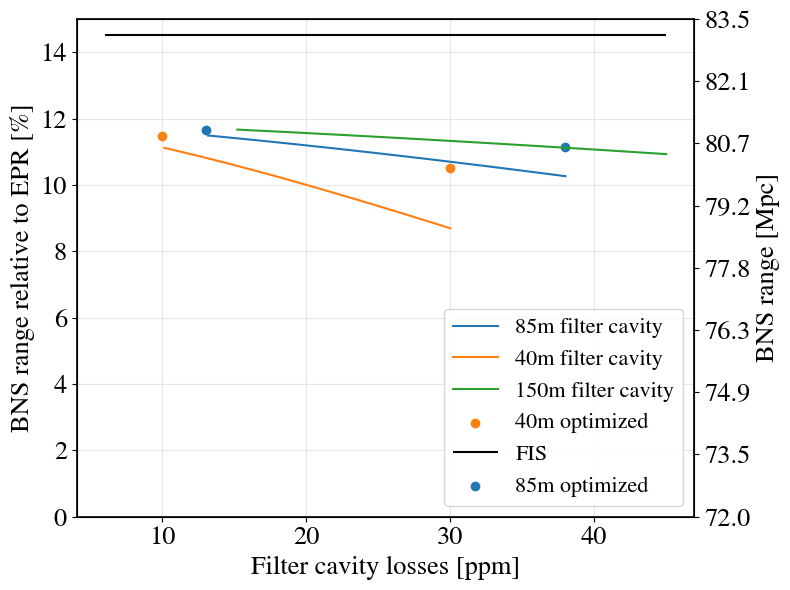}
    \caption{Detection range for KAGRA post-O5 after considering both quantum and classical noise (low quality factor suspension). The 85\,m space is current most allowed space in the KAGRA tunnel. All filter cavity losses value used here is between the best achievable value to the empirical line, as shown between the blue and orange line in Fig.\,\ref{fig:losses}. The optimization is done for the cases of 40\,m and 85\,m filter cavities for the lowest and highest filter cavity losses values.}
    \label{fig:range_compare_LQS}
\end{figure}

\begin{acknowledgments}
The authors thank Chan Park, Ray-Kuang Lee and the SQZ working group of KAGRA collaboration for helpful comments during the preparation of this manuscript. Y. Z. acknowledges support from the Joint Fund of Henan Province Science and Technology R\&D Program No. 235200810057 and the Henan Province High-Level Talent International Training Program. Z.-H. Z. acknowledges support from the National Natural Science Foundation of China under Grants Nos. 12433001 and 12021003. M. E. acknowledges support from the JSPS Grant-in-Aid for Scientific Research (Grants No.\,24K00649). This work is partially supported by the collaborative research program of the Institute for Cosmic Ray Research (ICRR) at the University of Tokyo. 

\end{acknowledgments}

\bibliography{references}

\end{document}